\documentclass[final]{svjour2}
\usepackage{graphicx}
\usepackage{rotating}
\usepackage{amssymb}
\usepackage{bm}
\usepackage{color}
\usepackage{mathptmx}
\usepackage{amsmath}
\usepackage{amsfonts}
\makeatletter
\journalname{Journal of Low Temperature Physics}
\begin{document}

\newcommand{\hdblarrow}{H\makebox[0.9ex][l]{$\downdownarrows$}-}
\title{Precursor phenomena of nucleations of quantized vortices in the presence of a uniformly moving obstacle in Bose-Einstein condensates}

\author{Masaya Kunimi$^1$ \and Yusuke Kato$^2$}

\institute{Department of Basic Science, The University of Tokyo.\\ Tokyo 153-8902, Japan\\
%Tel.:\\ Fax:\\
1:\email{kunimi@vortex.c.u-tokyo.ac.jp}
\\2:\email{yusuke@phys.c.u-tokyo.ac.jp}
}

%\date{XX.XX.2007}

\maketitle


%%%%%%%%%%%%%%%%%%%%%%%%%%%%%%%%%%%%%%%%%%%%%%%%%%%%%%%%%%%%%%

\begin{abstract}
We investigate excitations and fluctuations of Bose-Einstein condensates in a two-dimensional torus with a uniformly moving Gaussian potential by solving the Gross-Pitaevskii equation and the Bogoliubov equation. The energy gap $\Delta$ between the current-flowing metastable state (that reduces to the ground state for sufficiently slowly-moving potential) and the first excited state vanishes when the moving velocity $v$ of the potential approaches a critical velocity $v_{\rm c}(>0)$. We find a scaling law $\Delta \propto (1-|v|/v_{\rm c})^{\frac14}$, which implies that a characteristic time scale diverges toward the critical velocity. Near the critical velocity, we show that low-energy local density fluctuations are enhanced. These behaviors can be regarded as precursor phenomena of the vortex nucleation.

\end{abstract}
%%%%%%%%%%%%%%%%%%%%%%%%%%%%%%%%%%%%%%%%%%%%%%%%%%%%%%%%%%%%%%

%%%%%%%%%%%%%%%%%%%%%%%%%%%%%%%%%%%%%%%%%%%%%%%%%%%%%%%%%%%%%%
\section{Introduction}
%%%%%%%%%%%%%%%%%%%%%%%%%%%%%%%%%%%%%%%%%%%%%%%%%%%%%%%%%%%%%%

In cold atomic gases experiments of Bose-Einstein condensates(BECs) with blue-detuned laser beams, the critical velocity of superfluidity has been observed in simply-connected\cite{Raman1999,Onofrio2000,Inouye2001,Neely2010} and multiply-connected\cite{Ramanathan2011,Moulder2012,Wright2013} systems. In both cases, nucleations of the quantized vortices have been observed above the critical velocity. The observed critical velocity in the presence of obstacles is smaller than that of Landau's prediction\cite{Landau1941}.

Our goal is to reveal a mechanism of the breakdown of superfluidity due to the vortex nucleation. As a step toward this aim, we seek for signals of instability in spectral properties and fluctuations in the present paper. 
Theoretical studies of breakdown of superfluidity have been done by the Gross-Pitaevskii(GP) equation\cite{Gross_Pitaevskii} in various setups\cite{Frisch1992,Hakim1997,Jackson1998,Josserand1999,Winiecki1999,Jackson2000,Stiesberger2000,Huepe2000,Pavloff2002,Pham2002,Aftalion2003,Astrakharchik2004,El2006,Piazza2009,Sasaki2010,Kato2010,Fujimoto2011,Aioi2011,Dubessy2012,Piazza2013}. These studies have reported many nontrivial phenomena such as nonlinear dynamics of solitons or vortices etc. However, the properties of excitation spectrum related to the vortex nucleation in systems without translational or rotational symmetry have not been studied so far. An exception is a study on soliton nucleation in infinite 1D system with a point-like defect or higher dimensional system with a sheet-like potential\cite{Kato2010}. This work showed that dynamical local density fluctuations are enhanced near the critical velocity, as a precursor effect of breakdown of superfluidity. Since both solitons and quantized vortices are topological defects that kill superfluidity, we can expect that dynamical density fluctuations are enhanced near the critical velocity as a precursor effect of nucleation of vortices.

In this paper, we study the stability of a BEC confined in a two-dimensional torus with a moving Gaussian potential by the GP and the Bogoliubov equation\cite{Bogoliubov1947}. Our setup corresponds to a simplified model of the recent experiment of a BEC in a ring trap with a rotating weak link\cite{Wright2013}. Since we focus on the effect of the Gaussian potential, we neglect the surface and the centrifugal effects.

%%%%%%%%%%%%%%%%%%%%%%%%%%%%%%%%%%%%%%%%%%%%%%%%%%%%%%%%%%%%%%
\section{Model}\label{eq:sec:Model}
%%%%%%%%%%%%%%%%%%%%%%%%%%%%%%%%%%%%%%%%%%%%%%%%%%%%%%%%%%%%%%

We consider an $N$-Boson system confined in a two-dimensional torus $[-L/2, L/2)\times [-L/2, L/2)$ with a moving Gaussian potential $U(\bm{r}+\bm{v}t)$, where the velocity of the potential $-\bm{v}$ with $\bm{v}\equiv v\bm{e}_x (v>0)$ is anti-parallel to the unit vector $\bm{e}_x$ of $x$-direction. In the mean-field approximation, the GP equation\cite{Gross_Pitaevskii} in the moving frame with the potential is given by
\begin{eqnarray}
-\frac{\hbar^2}{2m}\nabla^2\Psi(\bm{r})+U(\bm{r})\Psi(\bm{r})+g|\Psi(\bm{r})|^2\Psi(\bm{r})=\mu\Psi(\bm{r}),\label{eq:GP_equation_in_the_moving_frame}
\end{eqnarray}
where $\Psi(\bm{r})$ is the condensate wave function(the order parameter of BEC), $U(\bm{r})\equiv U_0\exp{[-\left(\bm{r}/d\right)^2]}$ is the Gaussian potential, $g(>0)$ is the strength of the repulsive interaction, and $\mu$ denotes the chemical potential that is determined by the condition $N=\int d\bm{r}|\Psi(\bm{r})|^2$. The twisted periodic boundary condition in the moving frame is given by
\begin{eqnarray}
\Psi(\bm{r}+L\bm{e}_x)&=&e^{im vL/\hbar}\Psi(\bm{r}),\; \Psi(\bm{r}+L\bm{e}_y)=\Psi(\bm{r}),\label{eq:twisted_periodic_boundary_condition}
\end{eqnarray}
where $\bm{e}_y$ denotes the unit vector of $y$-direction. We note that the boundary condition (\ref{eq:twisted_periodic_boundary_condition}) does not change under the transformation $v\to v+2\pi\hbar n/(m L)$, where $n$ is an arbitrary integer. In the present work, we focus on the branch of the stationary flow states that reduces to the ground state when $|v|<\pi\hbar /(m L)$.

The Bogoliubov equation\cite{Bogoliubov1947} is given by
\begin{eqnarray}
\begin{bmatrix}
\mathcal{L} & -g[\Psi(\bm{r})]^2 \\
 g[\Psi^{\ast}(\bm{r})]^2  &-\mathcal{L}
\end{bmatrix}
\begin{bmatrix}
u_i(\bm{r}) \\
v_i(\bm{r})
\end{bmatrix}
=\epsilon_i
\begin{bmatrix}
u_i(\bm{r}) \\
v_i(\bm{r})
\end{bmatrix}
,\\
\mathcal{L}\equiv -\frac{\hbar^2}{2m}\nabla^2+U(\bm{r})-\mu+2g|\Psi(\bm{r})|^2,
\end{eqnarray}
where $u_i(\bm{r})$ and $v_i(\bm{r})$ are the wave functions of the excited state $i$ and $\epsilon_i$ is the excitation energy of the state $i$. The wave functions of the excited states satisfy the following orthonormal conditions $\int d\bm{r}[u_i(\bm{r})u_j^{\ast}(\bm{r})-v_i(\bm{r})v_j^{\ast}(\bm{r})]=\delta_{i j}$ and $\int d\bm{r}\left[u_i(\bm{r})v_j(\bm{r})-v_i(\bm{r})u_j(\bm{r})\right]=0$ and the boundary conditions $u_i(\bm{r}+L\bm{e}_x)=e^{imvL/\hbar}u_i(\bm{r}),\;u_i(\bm{r}+L\bm{e}_y)=u_i(\bm{r}), v_i(\bm{r}+L\bm{e}_x)=e^{-imvL/\hbar}v_i(\bm{r})$, and $v_i(\bm{r}+L\bm{e}_y)=v_i(\bm{r})$.

We take the unit of the length, velocity, and energy, respectively, as the healing length $\xi\equiv \hbar/\sqrt{m g n_0}$, the sound velocity $v_{\rm s}\equiv \sqrt{g n_0/m}$, and $\epsilon_0\equiv \hbar^2/(m\xi^2)=g n_0$, where $n_0\equiv N/L^2$ represents a mean-particle density. Here, we set $1/\sqrt{n_0\xi^2}=0.1$.

Our numerical calculations are performed in the $k$-space, that is, we expand $\Psi(\bm{r})e^{-imvx/\hbar}, u_i(\bm{r})e^{-imvx/\hbar}$, and $v_i(\bm{r})e^{imvx/\hbar}$ with the plane wave $e^{i\bm{k}\cdot\bm{r}}$, where $\bm{k}\equiv (2\pi/L)(n_x\bm{e}_x+n_y\bm{e}_y)$ and $n_x$ and $n_y$ are integers. Rewriting the GP and the Bogoliubov equations in terms of the expansion coefficients, we solve them numerically.  The solutions of the GP equation (\ref{eq:GP_equation_in_the_moving_frame}) are obtained by the imaginary time evolution method. The Bogoliubov equation is diagonalized by the same technique as that used in Ref.~\cite{Kunimi2012}. We have found that the dependence of the present results on the number of the basis is negligibly small\cite{Note1}.

%%%%%%%%%%%%%%%%%%%%%%%%%%%%%%%%%%%%%%%%%%%%%%%%%%%%%%%%%%%%%%
\section{Results}\label{sec:Results}
%%%%%%%%%%%%%%%%%%%%%%%%%%%%%%%%%%%%%%%%%%%%%%%%%%%%%%%%%%%%%%

\begin{figure}[t]
\centering
\includegraphics[width=11.5cm,clip]{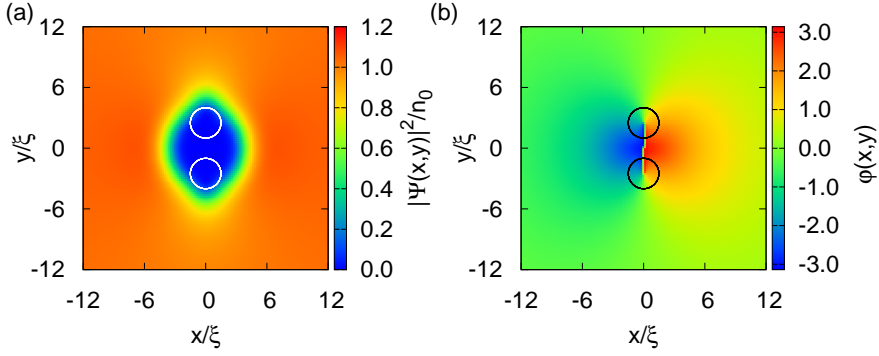}
\caption{(Color online) (a) Density profile and (b) phase profile except for the plane wave component $e^{i m v x/\hbar}$ for $L=48\xi$, $U_0=5\epsilon_0$, $d=2.5\xi$, $v=0.426545v_{\rm s}$. White and black circles represent the positions of the GVP.}
\label{fig:density_and_phase}
\end{figure}

First, we present the stable stationary solution of the GP equation near the critical velocity in Fig.~\ref{fig:density_and_phase}. Here, the critical velocity is defined as the modulus of the velocity far away from the obstacle, above which the stationary flow state becomes unstable. In the presence of the strong potential, we find that a bounded pair of vortices appears in the low density region. This vortex-pair is called ghost vortex pair(GVP)\cite{Kasamatsu2003,Fujimoto2011}. Although GVP exists, the excitation spectra (shown later) do not have anomaly and thus the stationary solution is stable or metastable. Other physical quantities except local phase spectral function (discussed later) do not have singularities either. The reason why the stationary solution is metastable can be understood that the GVP is pinned to the potential and free vortex motions that  cause the energy dissipation do not occur. In other parameter regions, particularly in weak potential strength cases(for example, $U_0=\epsilon_0$ and $d=2.5\xi$), the GVP does not appear in our numerical calculations. We typically investigate the range of the velocity for $10^{-6}\lesssim |(v_{\rm c}-v)/v_{\rm c}|\le 1$, where $v_{\rm c}$ is the critical velocity. We conclude that the GVP does not exist in the weak potential case within this range of the velocity.

\begin{figure}[t]
\centering
\includegraphics[width=11.0cm,clip]{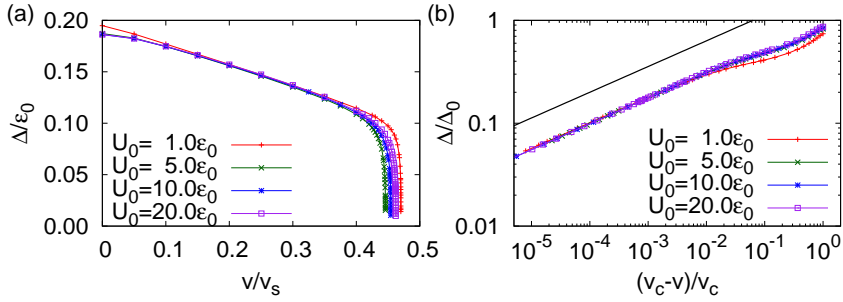}
\caption{(Color online) (a) Velocity dependence of the energy gap for $L=32\xi$ and $d=2.5\xi$. (b) Fitting result of (a). Solid black line in (b) represents the line $[(v_{\rm c}-v)/v_{\rm c}]^{1/4}$.}
\label{fig:energy_gap}
\end{figure}

Now we focus on the minimum eigenenergy ${\rm min}_i (\epsilon_i)\equiv \Delta$ (which we call the energy gap) of the Bogoliubov equation in finite size systems. We show the velocity dependence of the energy gap in Fig.~\ref{fig:energy_gap}(a), where we can see two features. One is the linearly decreasing of $\Delta$ as a function of $v$ in the small $v$ region. This behavior reflects the property of the energy gap $\Delta_{\rm uni}=2\pi g n_0/(L/\xi)[-v/v_{\rm s}+\sqrt{\pi^2/(L/\xi)^2+1}]$ in the uniform system with the shape of two-dimensional torus. Another feature we find in Fig.~\ref{fig:energy_gap}(a) is the steep decrease near $v=v_{\rm c}$. We assume a form  $\Delta=\Delta_0\left[(v_{\rm c}-v)/v_{\rm c}\right]^c$ with adjustable parameters $\Delta_0, v_{\rm c}$, and $c$ to fit the four sets of data shown in Fig.~\ref{fig:energy_gap}(a) near the critical velocity. Figure ~\ref{fig:energy_gap}(b) shows that the data points near the critical velocity collapse onto a single power law as a function of $(v_{\rm c}-v)/v_{\rm c}$ with the exponent $c\simeq 0.25$, which is nearly independent of the potential height, width, and the system size within our calculations(see Table.~\ref{tab:exponent}). 

Our results in finite size systems indicate that the energy gap vanishes at $v=v_{\rm c}$. From the viewpoint of non-linear physics, the origin of this behavior can be understood by the bifurcation of the solution of the non-linear differential equation. According to the bifurcation theory\cite{Guckenheimer1983}, eigenvalues of a linearized equation of an original non-linear differential equation become zero at the bifurcation point. Here, the original equation, the linearized equation, and the bifurcation point correspond to the GP equation, the Bogoliubov equation, and the critical velocity, respectively. Additionally, the eigenvalues obey a scaling law near the bifurcation point, which reflects the type of bifurcation. The power law behavior($c\simeq 0.25$) in the present system is consistent with a Hamiltonian saddle node bifurcation reported in Ref.~\cite{Huepe2000}. Physically, the existence of the dynamical scaling law of the energy gap implies that the characteristic time scale diverges toward the critical velocity.

So far, we revealed the characteristic properties of the excitation spectra in a non-uniform finite size system. In order to investigate further the instability, we calculate fluctuations from the wave functions of the excited states.  In the case of the soliton nucleation, the authors of \cite{Kato2010} introduced a local density spectral function $I_n(\bm{r}, \epsilon)$. In the Bogoliubov approximation, $I_n(\bm{r}, \epsilon)$ is given by
\begin{eqnarray}
I_n(\bm{r},\epsilon)&\equiv &\sum_i|\delta n_i(\bm{r})|^2\delta(\epsilon-\epsilon_i),\label{eq:local_spectral_function_definition}\\
\delta n_i(\bm{r})&\equiv&\Psi^{\ast}(\bm{r})u_i(\bm{r})-\Psi(\bm{r})v_i(\bm{r}).\label{eq:matrix_element_density_fluctuation}
\end{eqnarray}
According to Ref.~\cite{Kato2010}, the low-energy density fluctuations increase toward the critical velocity around the obstacle. 

\begin{figure*}[t]
\includegraphics[width=12.0cm,clip]{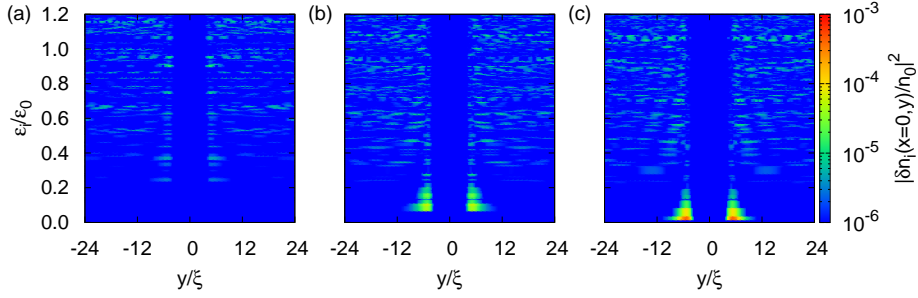}
\caption{(Color online) The matrix element of density fluctuation for $L=48\xi, U_0=10\epsilon_0, d=2.5\xi, x=0$, and (a)$v=0.1v_{\rm s}$, (b)$0.42v_{\rm s}$ and (c)$0.43033v_{\rm s}$, respectively.}
\label{fig:density_flctuation_48_10_25}
\end{figure*}

In our system, it is difficult to calculate the spectral function because of the discretized eigenvalues in finite size systems. Instead of calculating the spectral function, we plot the matrix element (\ref{eq:matrix_element_density_fluctuation}) in Fig.~\ref{fig:density_flctuation_48_10_25}. In the small velocity regime, the low-energy density fluctuation is small because of the existence of the energy gap(see Fig.~\ref{fig:density_flctuation_48_10_25}(a)). The low-energy local density fluctuations grow  with increasing the velocity shown in Figs.~\ref{fig:density_flctuation_48_10_25}(b) and (c). Combining the scaling law of the energy gap and the behavior of the dynamical density fluctuations, we can expect that the zero-energy eigenstate that contributes the density fluctuations exists at $v=v_{\rm c}$. In the previous work for the soliton nucleation in infinite 1D system\cite{Takahashi2009}, the zero-energy eigenstate related to the density fluctuations was obtained by analytical calculations and plays a crucial role of the breakdown of superfluidity. Therefore, in the case of the vortex nucleation, the density fluctuation is also expected to be a key to understand the breakdown of superfluidity.

Besides the local density spectral function (Eq.~(\ref{eq:local_spectral_function_definition})), the authors of \cite{Watabe2013} also introduced the local phase spectral function. In the Bogoliubov approximation, this is given by
\begin{eqnarray}
I_{\varphi}(\bm{r}, \epsilon)&=&\sum_i|\delta\varphi_i(\bm{r})|^2\delta(\epsilon-\epsilon_i),\;\delta\varphi_i(\bm{r})\equiv \frac{\Psi^{\ast}(\bm{r})u_i(\bm{r})+\Psi(\bm{r})v_i(\bm{r})}{2|\Psi(\bm{r})|^2}.\label{eq:matrix_element_phase_fluctuation}
\end{eqnarray}
They showed that the local phase spectral function is not useful to detect the critical velocity. The local phase fluctuation is nearly independent of the velocity\cite{Watabe_private} because the zero mode related to the phase fluctuations always exists\cite{note1} as a result of the $U(1)$ symmetry breaking. Furthermore, there is no phase singularity in 1D system. In our system, however, the matrix element of the phase fluctuation (\ref{eq:matrix_element_phase_fluctuation}) diverges locally because of the presence of the GVP. Although we can detect the existence of the GVP from the divergence of the matrix element of the local phase fluctuation, the critical velocity can not be detected from it because the GVP appears for $v<v_{\rm c}$. 

Finally, we make a remark on dynamics. 
We have checked that the vortex pair is always nucleated above the critical velocity in the earlier stage of a real-time dynamics in our system, regardless of the presence/absence of GVP in the initial states. 

On the other hand, the low energy density fluctuations are enhanced regardless of existence/absence of the GVP. Therefore, the enhancement of low energy density fluctuations can be regarded as a precursor phenomenon of nucleations of quantized vortices.

\begin{table}[h]
\centering
\caption{Fitting results.}
\begin{minipage}[h]{0.47\linewidth}
\begin{tabular}{ccccc|c|c|c|c}\hline
$L/\xi$&$U_0/\epsilon_0$&$d/\xi$& exponent \\ \hline
18 &  5 & 0.5 &0.24814(6) \\
24 & 0.5  & 2.5 & 0.24842(7) \\
24 & 10 & 2.5 & 0.25133(8) \\
32 & 1 &2.0& 0.24879(6) \\ %\hline
32 & 1 &2.5& 0.24880(5) \\ %\hline
32 & 1 &5.0& 0.25191(5)  \\ %\hline
32 & 5 &2.0& 0.24849(6)  \\ \hline
%&&&&&&&&\\ \hline
\end{tabular}
\end{minipage}
\begin{minipage}[h]{0.47\linewidth}
\begin{tabular}{ccccc|c|c|c|c}\hline
$L/\xi$&$U_0/\epsilon_0$&$d/\xi$& exponent \\ \hline
32 & 5 &2.5& 0.25125(3)  \\ %\hline
32 & 10 &2.0& 0.24905(9)  \\ %\hline
32 & 10 &2.5& 0.25139(7)  \\ %\hline
32 & 20 &2.0& 0.24994(5)  \\ %\hline
32 & 20 &2.5& 0.25130(6)  \\ %\hline
48 & 10 &2.5& 0.24875(9)  \\ %\hline
64 & 10 &2.5& 0.24862(6)  \\ \hline
%&&&&&&&&\\ \hline
\end{tabular}
\end{minipage}
\label{tab:exponent}
\end{table}

%%%%%%%%%%%%%%%%%%%%%%%%%%%%%%%%%%%%%%%%%%%%%%%%%%%%%%%%%%%%%%
\section{Summary and discussion}\label{sec:summary_and_discussion}
%%%%%%%%%%%%%%%%%%%%%%%%%%%%%%%%%%%%%%%%%%%%%%%%%%%%%%%%%%%%%%
In summary, we studied the excitation spectra and the fluctuations of a BEC in a 2D torus with a uniformly moving Gaussian potential by solving the GP and the Bogoliubov equation. We found that the first excited energy(or the energy gap) obeys a scaling law, which implies a diverging time scale near the critical velocity. Further we found enhancement of dynamical low-energy local density fluctuation near the critical velocity. These two behaviors can be regarded as precursor effects of the vortex nucleation.

A future work is to elucidate the relation between enhancement of dynamical density fluctuations and the breakdown of superfluidity(or energy dissipation). 
It is difficult to measure the local density spectral function experimentally. We thus must find the relation between the local density spectral function and other physical quantities such as transport coefficients. In the case of the Landau instability, the authors of Ref.~\cite{Astrakharchik2004} showed that the energy dissipation and the density fluctuation are related through the drag force by using the perturbative approach. Although their approach is not applicable for the vortex and soliton the nucleation as it stands, their results give us a clue to the relation between the vortex nucleation and the energy dissipation.

%%%%%%%%%%%%%%%%%%%%%%%%%%%%%%%%%%%%%%%%%%%%%%%%%%%%%%%%%%%%%%
\begin{acknowledgements}

We would like to thank S. Watabe and I. Danshita for fruitful discussions and D. Yamamoto for useful comments. M. K. acknowledges the support of a Grant-in-Aid for JSPS Fellows (239376).
This work is supported by KAKENHI (21540352) and (24543061) from JSPS and (20029007) from MEXT in Japan. 

\end{acknowledgements}

%%%%%%%%%%%%%%%%%%%%%%%%%%%%%%%%%%%%%%%%%%%%%%%%%%%%%%%%%%%%%%

%\pagebreak
%%%%%%%%%%%%%%%%%%%%%%%%%%%%%%%%%%%%%%%%%%%%%%%%%%%%%%%%%%%%%%


\begin{thebibliography}{99}
\bibitem{note1}
$u(\bm{r})=\Psi(\bm{r})$ and $v(\bm{r})=\Psi^{\ast}(\bm{r})$ are always solutions of the Bogoliubov equation with a zero eigenvalue.
\end{thebibliography}
\end{document}